\documentclass{emulateapj}
\usepackage{epsfig}
\usepackage{apjfonts}
\usepackage{graphics}

\newcommand{\ssim}{{\cal S}}
\newcommand{\tsim}{{\cal T}}
\newcommand{\HI}{\ion{H}{1}}

\newcommand{\CIII}{\ion{C}{3}} 
 
\newcommand{\CIV}{\ion{C}{4}} 
\newcommand{\SiIII}{\ion{Si}{3}} 
 
\newcommand{\SiIV}{\ion{Si}{4}} 
 
\newcommand{\OVI}{\ion{O}{6}}

\lefthead{Aguirre et al.}
\righthead{Simulations and observations of the IGM}

\def\gsim{\;\rlap{\lower 2.5pt
 \hbox{$\sim$}}\raise 1.5pt\hbox{$>$}\;}
\def\lsim{\;\rlap{\lower 2.5pt
   \hbox{$\sim$}}\raise 1.5pt\hbox{$<$}\;}
\def\msol{{\rm\,M_\odot}}

\def\spose#1{\hbox to 0pt{#1\hss}}

\def\zsol{{\,Z_\odot}}

\shorttitle{Simulations vs. observed intergalactic metals}
\shortauthors{Aguirre et al.}

\begin{document}
	
\title{Confronting cosmological simulations with observations of
  intergalactic metals}
\author{Anthony Aguirre\altaffilmark{1}, Joop~Schaye\altaffilmark{2}, \\
Lars Hernquist\altaffilmark{3}, Scott Kay\altaffilmark{4,5}, Volker Springel\altaffilmark{6}, Tom Theuns\altaffilmark{7,8}}
\altaffiltext{1}{Department of Physics, University of California at Santa Cruz,1156 High Street, Santa Cruz, CA  95064; aguirre@scipp.ucsc.edu}
\altaffiltext{2}{School of Natural Sciences, Institute for Advanced
Study, Einstein Drive, Princeton NJ 08540}
\altaffiltext{3}{Harvard-Smithsonian Center for Astrophysics, 60 Garden Street, Cambridge MA 02138}
\altaffiltext{4}{Astronomy Centre, University of Sussex, Falmer, Brighton BN1 9QH, UK}

\altaffiltext{5}{Astrophysics, Denys Wilkinson Building, Oxford OX1 3RH, UK}

\altaffiltext{6}{Max Planck Institute for Astrophysics, Karl-Schwartzschild Strasse 1, Garching, Munich, D-85740, Germany}
\altaffiltext{7}{Institute for Computational Cosmology, Department of
Physics, University of Durham, South Road, Durham, DH1 3LE, UK}
\altaffiltext{8}{University of Antwerp, Campus Drie Eiken,
Universiteitsplein 1, B-2610 Antwerp, Belgium}

\setcounter{footnote}{0}

\begin{abstract}
Using the statistics of pixel optical depths, we compare \HI, \CIV\
and \CIII\ absorption in a set of six high quality $z\sim 3-4$ quasar
absorption spectra to that in spectra drawn from two different
state-of-the-art cosmological simulations that include galactic
outflows. We find that the simulations predict far too little \CIV\
absorption unless the UVB is extremely soft, and always predict far
too small \CIII/\CIV\ ratios.  We note, however, that much of the
enriched gas is in a phase ($T\sim 10^5-10^7\,$K,
$\rho/\langle\rho\rangle\sim 0.1-10$, $Z\gtrsim 0.1 Z_\odot$) that
should cool by metal line emission -- which was not included in our
simulations. When the effect of cooling is modeled, the predicted
\CIV\ absorption increases substantially, but the \CIII/\CIV\ ratios
are still far too small because the density of the enriched gas is too
low. Finally, we find that the predicted metal distribution is much
too inhomogeneous to reproduce the observed probability distribution
of \CIV\ absorption.  These findings suggest that strong $z\lesssim 6$
winds cannot fully explain the observed enrichment, and that an
additional (perhaps higher-$z$) contribution is required.
\end{abstract}

\section{Introduction}
\label{sec-intro}

Analysis of quasar absorption spectra has revealed that the
intergalactic medium (IGM) has been polluted with heavy elements such
as carbon, silicon and oxygen (see, e.g., Cowie et al.\ 1995; Songaila
\& Cowie 1996; Ellison et al.\ 2000; Schaye et al. 2003, hereafter
S03; Aguirre et al. 2004, hereafter A04; Simcoe et al 2004; Aracil et
al. 2004; Boksenberg et al. 2003) At the same time, observations of
starburst galaxies (e.g., Shapley et al.\ 2003) have revealed powerful
galactic outflows resulting from feedback processes in galaxies with
rapid star formation.

This has led to a picture in which the observed enrichment results
from a phase of strong galactic outflows at $z \gtrsim 2$ during the
epoch of galaxy formation, and various numerical (e.g., Gnedin 1998;
Aguirre et al. 2001; Scannapieco, Ferrara, \& Madau 2002; Cen,
Nagamine \& Ostriker 2004) and semi-analytic (e.g., Furlanetto \& Loeb
2003) models have been able to very roughly account for the observed
level of metal enrichment.  But these comparisons have left many
unanswered questions: can, for example, simulations reproduce the
detailed density- and redshift-dependent metal distribution? Can they
reproduce the abundance ratios of different elements and ions? How is
enrichment tied to feedback in galaxies, which is required to suppress
runaway star formation?

Recently, both the numerical simulations and observational analyses
have improved to the degree that some of these questions can be
meaningfully addressed. On the observational side, statistical
analyses of pixel optical depths (S03; A04) have inferred the
distribution of carbon and silicon using absorption by \HI, \CIV,
\CIII, \SiIV, and \SiIII\ (see also Simcoe et al. 2004, who finds
consistent results from line fitting of \CIV\ and \OVI.)  These
studies have shown quantitatively that the observed intergalactic
carbon enrichment is highly inhomogeneous, density-dependent, nearly
redshift-independent, underabundant (relative to silicon), relatively
cool ($T < 10^5\,$K), and persistent at some level even in gas near
the cosmic mean density.  Meanwhile, hydrodynamic simulations (e.g.,
Theuns et al. 2002, hereafter T02; Springel \& Hernquist 2003a,
hereafter SH03) have been produced that probably include all of the
galaxies relevant for $z \lesssim 6$ enrichment, use prescriptions for
feedback that generate galactic winds, and track metals. SH03 and
Hernquist \& Springel (2003) have shown in detail that the star
formation rate of their simulations has converged, i.e. that the
included feedback is sufficient to solve the overcooling problem; T02
have shown that their simulation does {\em not} significantly disrupt
the Ly$\alpha$ forest, but can roughly account for earlier
observations of \CIV\ absorption if the UVB is extremely soft and the
yield is $3\times$ solar.

Clearly, it is of interest to carry out detailed comparisons between
state-of-the art observations and simulations that include galactic
winds.  One way to do this is to simply compare the observationally
inferred distribution of metals as a function of density and redshift
to that predicted by the simulations.  This could, however, be
misleading. For example, hot, collisionally ionized carbon would be
undetectable by observational studies focusing on \CIV. A far more
direct and robust method is to directly compare simulated and observed
absorption spectra. This Letter describes such a comparison.  We will
draw several qualitatively new conclusions from a comparison of
absorption by \HI, \CIV, and \CIII\ in in a set of 6 $z\sim 3$ QSO
spectra to that in simulated spectra drawn from the SH03 and T02
simulations.

\section{Observations, simulations, and method}
\label{sec-simobs}

We compare our simulations to the observed pixel statistics published
in S03 for the redshift range $2.479 \le z \le 4.033$ (the full sample
covers $1.654 \le z \le 4.451$; we employ a smaller range so that all
of the data can be combined without binning in redshift). The data
come from six quasar spectra: Q0420-388, Q1425+604, Q2126-158,
Q1422+230, Q0055-269, and Q1055+461, that were taken with either the
Keck/HIRES or the VLT/UVES instrument. See Table 1 of S03 for
information on the sample.

The simulated spectra are drawn from cosmological SPH simulations,
using the method described in A02: noise, detector resolution,
wavelength coverage, and pixelization are chosen to match the
corresponding observed spectra. The ionization balance is computed
using CLOUDY\footnote{See
\texttt{http://www.pa.uky.edu/$\sim$gary/cloudy}.}, with the same
three models for the spectral shape of the UV background (UVB) as we
used in S03 and A04: `QG' is a Haardt \& Madau (2001) model with
contributions from quasars and galaxies; `Q' includes quasars only,
and `QGS' is a softened `QG': the flux is reduced by 90\% above 4 Ryd.
All models are normalized to the \HI\ ionization rate measured by S03,
and the simulation metallicities are converted to carbon number
densities using the solar abundance $({\rm C/H})_\odot=-3.45$ of
Anders \& Grevesse (1989).

Results are shown for three simulations.  The first, `NF', was used
and described in A02, S03 and A04: it uses $2\times 256^3$ particles
in a $12h^{-1}$\,Mpc box with $(\Omega_m, \Omega_\Lambda, \Omega_bh^2,
h, \sigma_8, n, Y) = (0.3, 0.7, 0.019, 0.65, 0.9, 1.0, 0.24)$. This
simulation has no galactic outflows, but for each UVB model we add to
the particles the carbon distribution inferred in S03 for that UVB
(see Tables 2 and 3 of S03). The second simulation, `\tsim', is
described in T02: it uses $2\times 128^3$ particles in a
$5h^{-1}$\,Mpc box with the same cosmological parameters as NF.  Here,
however, all supernova thermal energy is deposited as feedback, and
gas is prevented from cooling for $10^7\,$yr (see Kay et. al. 2002) so
that strong winds are generated which enrich the IGM. The third
simulation, `\ssim', is described in SH03 as their `Q4' model: it uses
$2\times 216^3$ particles in a $10h^{-1}\,$Mpc box with $(\Omega_m,
\Omega_\Lambda, \Omega_bh^2, h, \sigma_8, n, Y) = (0.3, 0.7, 0.02,
0.7, 0.9, 1.0, 0.24)$.  It employs both a sub-grid feedback
prescription and a wind mechanism in which a velocity of 484\,km/s is
imparted to certain gas particles in star-forming regions (see
Springel \& Hernquist 2003b for details).  The simulations have
approximately the same mass resolution (the baryon particle mass is
$\approx 1.1\times10^6\msol$ in both), which is adequate to resolve
the Ly$\alpha$ forest and to include relatively small galaxies. The
use of two simulations allows us to compare the effects of two
different feedback prescriptions.

To compare the carbon absorption in simulated and observed spectra, we
have employed the pixel optical depth technique described in A02, S03
and A04. This technique (see Cowie \& Songaila 1998; Ellison et
al. 2000; A02; S03) has several advantages over traditional
line-fitting. Perhaps most important here is the ability to measure
\CIII\ absorption, which is very difficult to do using line fitting
because \CIII\ is not a multiplet and falls in the Ly$\beta$
forest. Optical depths for \HI\,(Ly$\alpha$), \CIV\,(1548\AA), and
\CIII\, (977\AA)\, absorption are extracted from each spectrum for the
\HI\ absorption region between the QSO's Ly$\alpha$ and Ly$\beta$
emission wavelengths, excluding also a small region near the QSO to
avoid proximity effects (see S03).  The optical depths are corrected
for strong contaminating lines, \CIV\ self-contamination by its
doublet, and Ly$\alpha$ contamination of \CIII, as described in A02
and S03.

The pixel optical depths are binned for each QSO and the QSOs are
combined as described in A04. In brief, for each QSO the \CIV\ optical
depths $\tau_{\rm CIV}$ are binned in \HI\ optical depth $\tau_{\rm
HI}$ and the median is taken in each bin.  Next, the
noise/contamination level, i.e.\ the median $\tau_{\rm CIV}$ optical
depth at very low $\tau_{\rm HI}$, is subtracted from each bin and the
ratio $({\rm median}\, \tau_{\rm CIV})/\tau_{\rm HI}$ is computed.
Finally, $\tau_{\rm HI}$ bins for different QSOs are combined. The
result is a plot of median corrected $\tau_{\rm CIV}/\tau_{\rm HI}$
vs. $\tau_{\rm HI}$, as shown in Fig.~\ref{fig-civnocool} (upper
left).  The same procedure is applied to obtain plots of $\tau_{\rm
CIII}/\tau_{\rm CIV}$ vs. $\tau_{\rm CIV}$.

\section{Results}
\label{sec-res}

\begin{figure*}
\epsscale{1.1}
\plottwo{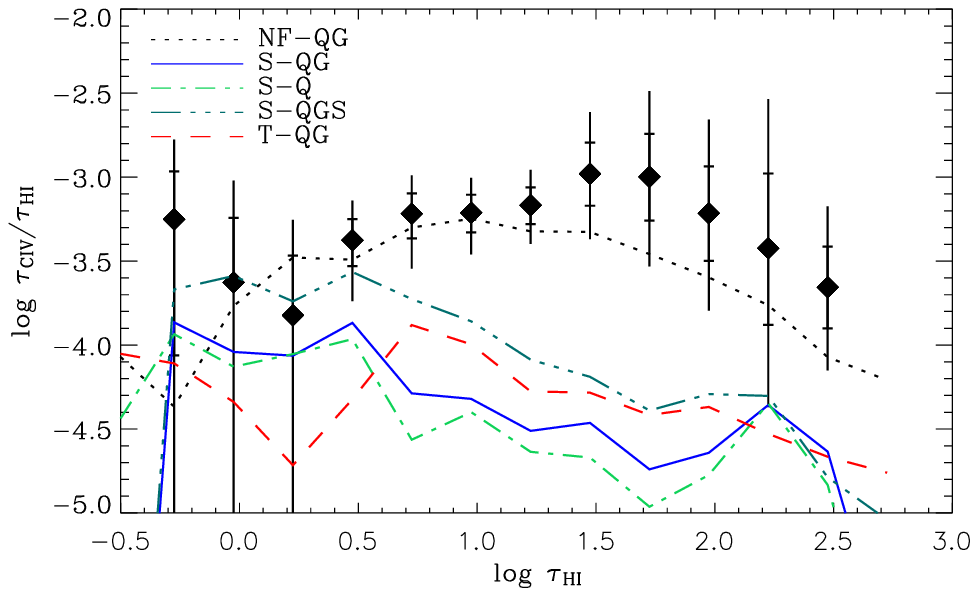}{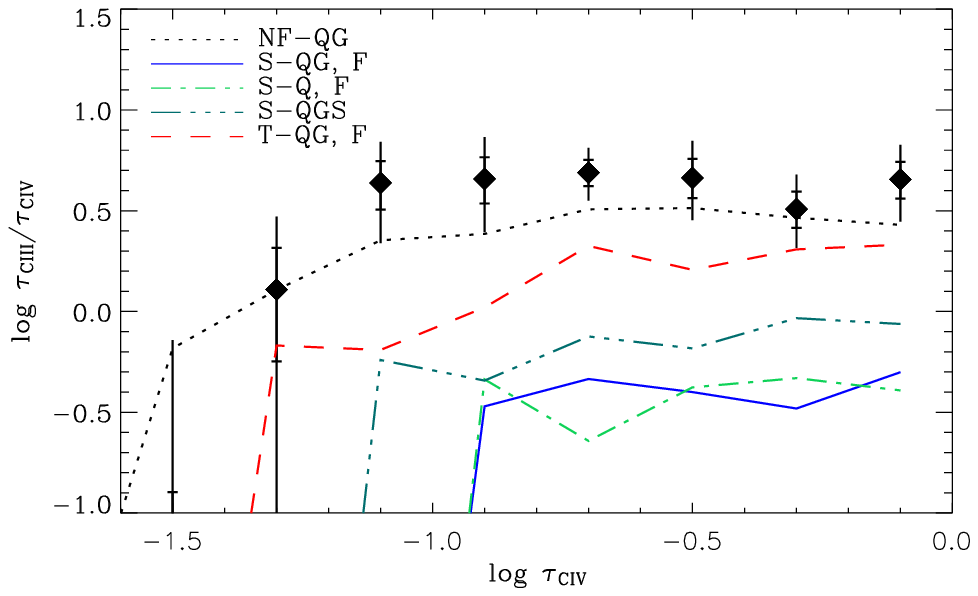}
\plottwo{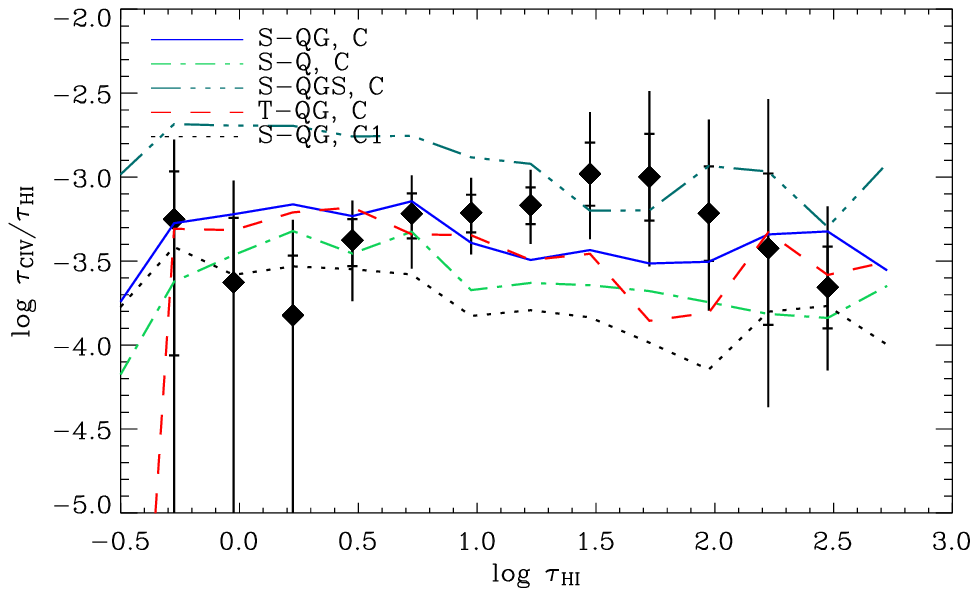}{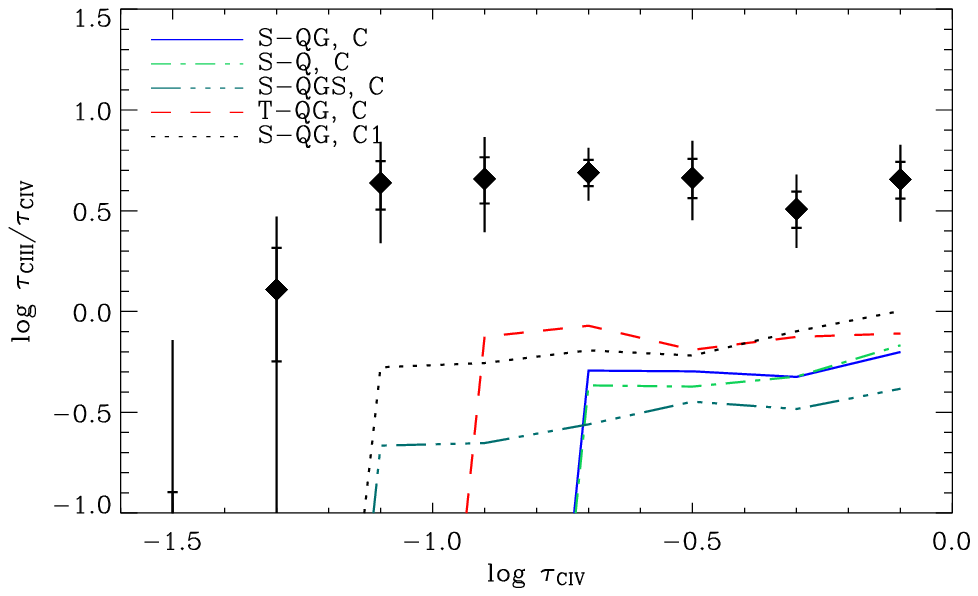}
\figcaption[] {Binned optical depth ratios $({\rm med}\,\tau_{\rm
CIV})/\tau_{\rm HI}$ vs.\ $\tau_{\rm HI}$ (\emph{left}) and $({\rm
med}\,\tau_{\rm CIII})/\tau_{\rm CIV}$ vs.\ $\tau_{\rm CIV}$
(\emph{right}) for observations (data points with 1- and 2-$\sigma$
errors) and various models. \emph{Top:} Models NF-QG (the Q and QGS
backgrounds yield similar results), \ssim-QG \ssim-Q, \ssim-QGS, and
\tsim-QG, as per the legend. The effect of changing the UVB in the
\tsim\ simulations is similar. \emph{Bottom:} as in the top panels,
but particles with cooling time $t_c < t_H$ are set to $T=2\times
10^4\,$K. The same models are shown, except that the dotted line
(`C1') corresponds to \ssim-QG if only particles with $t_c < 0.1 t_H$
cool (the effect of this change is similar for the other models).
\label{fig-civnocool}}
\end{figure*} 

Figure~\ref{fig-civnocool} shows median optical depths $({\rm
med}\,\tau_{\rm CIV})/\tau_{\rm HI}$ vs.\ $\tau_{\rm HI}$
(\emph{left}) and $({\rm med}\,\tau_{\rm CIII})/\tau_{\rm CIV}$ vs.\
$\tau_{\rm CIV}$ (\emph{right}) for observations (data points) and
three simulations.  The dotted line shows the non-feedback (NF)
simulation upon which the carbon distribution of S03 has been
imposed.\footnote{At high-$\tau_{\rm HI}$ the predicted $\tau_{\rm
CIV}/\tau_{\rm HI}$ (coming mostly from Q1422+230) is a bit low
because S03 forced a power-law fit to $Z(\delta)$ with a
redshift-independent index.} The other lines show the SH03 (\ssim) and
T02 (\tsim) simulations with the three choices of UVB (QG, Q, and
QGS). Except for the extremely soft UVB QGS, the simulations woefully
under-predict the median \CIV\ absorption at all $\tau_{\rm HI}$.  In
Table 1 (column 4) we quantify this by providing the best-fit offset
to each set of simulated \CIV/\HI\ optical depths (e.g., simulation
\ssim-QG is $\approx 1.2\,$dex too low). A super-solar yield could
somewhat ameliorate this but seems unlikely for carbon, which is
underabundant relative to $\alpha$-elements in the IGM (e.g., A04).
Although the QGS models are only 0.26-0.64 dex too low overall, they
cannot reproduce the observed shape of $\tau_{\rm CIV}/\tau_{\rm HI}$
vs.\ $\tau_{\rm HI}$: there is too much absorption at low density (low
$\tau_{\rm HI}$) and too little at high-density.

\begin{figure}
\epsscale{1.05} \plotone{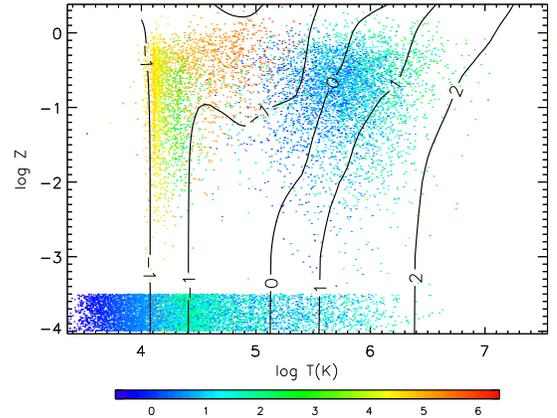} \figcaption[]{Metallicities (in solar
units) and temperatures of particles in the \ssim\ simulation (only a
random 0.5\% of metal enriched particles are shown, and 0.05\% of the
metal-free particles are shown in the scattered bar near
$Z=10^{-4}$). The particle colors reflect the log of the gas
overdensity as shown by the colored bar: intergalactic $\delta \alt
100$ gas is blue-green; galactic gas (with effective temperatures
given by the sub-grid multiphase model) is yellow-red.  The
contours indicate $\log(t_c/t_H)$ at $\delta=1$ and $z=3$.  Since $t_c
\propto \delta^{-1}$, particles inside the `0' contour at $\delta=1$
can cool, as can particles at $\delta=10$ inside the `1' contour.
\label{fig-zvt}}
\end{figure}

The prime reason for this failure can be seen in Fig.~\ref{fig-zvt},
which shows the metallicity, temperature, and density of a random
subsample of the \ssim\ simulation particles (the \tsim\ simulation is
similar).  The metal rich intergalactic (overdensity $\log\delta < 2$)
gas is almost entirely at $T=10^5-10^7\,$K.  Because the \CIV/C
fraction rapidly falls off at $T \gtrsim 10^5\,$K, this gas is
essentially invisible in \CIV, except at very low density if the UVB
is extremely soft.

Even for such an extreme UVB, however, the feedback simulations
predict far too little absorption by \CIII\ relative to \CIV, as can
be seen from the upper-right panel of Fig.~\ref{fig-civnocool} which
shows $({\rm med}\,\tau_{\rm CIII})/\tau_{\rm CIV}$ vs.\ $\tau_{\rm
CIV}$ for the same models\footnote{For the Q and QG UVBs, there is
insufficient CIV absorption to get a signal; we have thus run models
for higher yields (see Table 1).}. The \CIII/\CIV\ ratio drops rapidly
for $T \gtrsim 10^5\,$K and falls roughly linearly with decreasing
density for $T\sim 10^4\,$K (See Fig. 7 of S03). Thus, the fact that
the \CIV\ visible in the feedback simulations is accompanied by
insufficient \CIII\ means that the enriched gas is too hot and/or of
too low density. Note, on the other hand, that the NF simulation
reproduces the \CIII/\CIV\ values quite well once the carbon
distribution is chosen to match the \CIV/\HI\ values.

The problem that the metals in the feedback simulations are too hot
may, however, have a solution.  In both simulations the enriched gas
is relatively metal-rich by IGM standards ($Z \sim 0.1-1\zsol$) and
should, in fact, be able to cool via metal line emission, which was
{\em not} included in the simulations. The contours in
Fig.~\ref{fig-zvt} show $\log(t_{c}/t_H)$, where $t_c$ is the
radiative cooling time for gas that is in collisional ionization
equilibrium computed for overdensity $\delta\equiv
\rho/\langle\rho\rangle=1$ ($t_c \propto \delta^{-1}$) using
Sutherland \& Dopita (1993), and $t_H$ is the $z=3$ Hubble
time. Nearly all of the $\delta \sim 10$ gas and much of the
$\delta\sim 1$ gas has $t_{c} < t_H$ and should, in the absence of
heating, cool to $T\sim 10^4\,$K which would increase its visibility
in \CIV.  To test the importance of cooling, we have generated
simulated spectra for which all particles with $t_c < t_H$ are set to
$T=2\times 10^4\,$K; see bottom panels of Fig.~\ref{fig-civnocool}. In
this case, the QG simulations can roughly match the observed
$\tau_{\rm CIV}/\tau_{\rm HI}$ values, although the trend with
$\tau_{\rm HI}$ is still not reproduced. The QGS models with cooling
now predict too much \CIV\ absorption.

The employed cooling prescription is rather {\em ad hoc}. The particle
metallicities may be unreliable because the simulation assumes perfect
mixing at the particle level and zero mixing between particles once
they leave the star forming gas. Furthermore, heating is ignored in
the calculation of the cooling times, the gas may not be in
collisional ionization equilibrium, conduction is ignored, and the
density may not remain constant as the gas cools. To further
illustrate the sensitivity of the results to the cooling prescription,
the dotted curves in the lower panels of Fig.~\ref{fig-civnocool} show
the results for the S-QG simulation if the gas cools only when $t_c <
0.1t_H$. The \CIV\ absorption is lower by about 0.5~dex, nearly
independent of $\tau_{\rm HI}$. Varying the final temperature between
$1-3\times 10^4\,$K has a relatively small effect.

Although metal line cooling may help resolve the discrepancy for
\CIV/\HI, it actually makes the problem worse for \CIII/\CIV\ as can
be seen from the bottom right panel of Fig.~\ref{fig-civnocool}. This
indicates that in the simulations the \CIV\ absorption arises in gas
with too low a density. The level of $\log \tau_{\rm CIII}/\tau_{\rm
CIV}\sim -0.5$ predicted by the simulations indicates that the \CIV\
absorption with $\tau_{\rm CIV} \gtrsim 0.1$ arises in $\delta\sim
1-5$ gas, whereas the corresponding observed absorption appears to
occur in gas of $\delta\sim 10-30$ (see Fig.\ 7 of S03).  The increase
in density that would very likely accompany the gas cooling may
alleviate this problem, but this will have to be determined using
future simulations that include metal-line cooling.

Information on the homogeneity of the observed metal distribution can
be inferred by using additional percentiles of the $\tau_{\rm
CIV}(\tau_{\rm HI})$ distribution (S03).  In columns 3--5 of Table~1
we give the best-fit offset to the simulations in the 31st, 50th and
84th percentiles of the \CIV\ distribution.  We find that models that
can roughly reproduce the medians cannot simultaneously reproduce the
other percentiles. In all cases, the metals are {\em too
inhomogeneously distributed} as compared to the observations (e.g., in
S-QG with cooling, the 31st percentile is too low by $\sim0.8\,$dex
while the 84th is too high by $\sim0.6\,$dex, indicating a wider
distribution).  This appears to be true independent of the UVB and the
cooling prescription and is hence a rather robust inconsistency. It
is also interesting because while small-scale unresolved physical
effects such as a multiphase structure of the IGM and outflows (or
emission from interfaces between phases) are potentially quite
important, they seem likely to make the metals appear {\em less}
rather than more uniformly distributed.

%\begin{deluxetable*}{llccccll} 
\tabletypesize{\scriptsize}
\begin{deluxetable}{llcccc}
\tablecolumns{10} 
\tablewidth{0pc} 
\tablecaption{Comparisons made \label{tbl:comp}}
\tablehead{ Sim-
& & \multicolumn{3}{l}{Offset\tablenotemark{a} for percentile\tablenotemark{b} in CIV/HI} & in CIII/CIV\\
%\colhead{Offs.\tablenotemark{c}} & 
\colhead{UVB} & \colhead{Cool\tablenotemark{c}} & \colhead{31} & \colhead{50} &
 \colhead{84}
& \colhead{50}} 
\startdata %
NF-QG &  -  &  $  0.26 \pm 0.13 $  &  $  0.15 \pm 0.06 $ & $  0.12 \pm 0.05 $   &  $  0.16 \pm 0.05 $ \\
\ssim-QG &  -  &    NA\tablenotemark{d}   &  $  1.22 \pm 0.09 $  &  $  0.73 \pm 0.07 $  &  $  1.00 \pm 0.11$\tablenotemark{e}\\
\ssim-Q &  -  &  NA\tablenotemark{d}    &  $  1.41 \pm 0.10 $  &  $  0.73 \pm 0.07 $ & $  1.05 \pm 0.12 $\tablenotemark{e}\\
\ssim-QGS &  -  &    $  3.46 \pm  0.24 $& $  0.64 \pm 0.06 $ & $  0.41 \pm 0.05 $  &  $  0.73 \pm 0.08 $ \\
\tsim-QG &  -  &   NA\tablenotemark{d}  &  $  0.94 \pm 0.07 $  &  $  0.69 \pm 0.06 $   &  $  0.41 \pm 0.07 $\tablenotemark{e}\\
\tsim-Q &  -  &   NA\tablenotemark{d}  &  $  1.43 \pm 0.10 $   & $  0.94 \pm 0.07 $  &  $  0.58 \pm 0.07 $\tablenotemark{e}\\
\tsim-QGS &  -  &  $  1.40 \pm 0.22 $ &   $  0.26 \pm 0.06 $ & $  0.02 \pm 0.05 $ & $  0.56 \pm 0.07  $t \\
\ssim-QG &  C  & $  0.82 \pm 0.17 $   &  $  0.05 \pm 0.05 $   & $ -0.59 \pm 0.07  $   & $  0.76 \pm 0.07$  \\
\ssim-Q &  C  &   $  1.14 \pm 0.24 $   & $  0.28 \pm 0.06 $  & $ -0.16 \pm 0.07 $   &  $  0.87 \pm 0.07$ \\
\ssim-QGS &  C  &  $ -0.01 \pm 0.13 $    & $ -0.45 \pm 0.06 $  &  $ -1.42 \pm 0.05 $  & $  0.96 \pm 0.08 $  \\
\tsim-QG &  C  &  $  1.18 \pm 0.19 $    & $  0.13 \pm 0.05 $  & $ -0.31 \pm 0.06  $  & $  0.77 \pm 0.08 $  \\
\tsim-Q &  C  &  $  1.29 \pm 0.21 $   &  $  0.42 \pm 0.06 $  &   $  0.04 \pm 0.06 $ & $  0.68 \pm 0.08 $   \\
\tsim-QGS &  C  &  $  0.10 \pm 0.14 $    &  $ -0.28 \pm 0.06 $ &  $ -1.27 \pm 0.06 $  & $  1.07 \pm 0.12 $  \\
\ssim-QG &  C1  &   $  2.92 \pm 0.20 $  & $  0.48 \pm 0.06 $   &  $  0.11 \pm 0.06 $  &  $  0.65 \pm 0.06 $ \\
\ssim-Q &  C1  &  $  3.34 \pm 0.27 $  &  $   0.63 \pm 0.06 $   & $  0.34 \pm 0.06 $   &  $  0.78 \pm 0.07 $ \\
\ssim-QGS &  C1  & $  1.60 \pm 0.22 $   &  $  0.20 \pm 0.06 $   & $ -0.46 \pm 0.06 $   &  $  0.75 \pm 0.06 $ \\
\tsim-QG &  C1  &  $  1.86 \pm 0.23 $  & $  0.44 \pm 0.06 $   &  $  0.18 \pm 0.05 $  &  $  0.57 \pm 0.06 $ \\
\tsim-Q &  C1  &   $  4.65 \pm 0.32 $    & $  0.81 \pm 0.07 $  &  $  0.47 \pm 0.05 $   & $  0.44 \pm 0.07 $  \\
\tsim-QGS &  C1  &  $  0.68 \pm 0.18 $   & $  0.09 \pm 0.06 $  &  $ -0.42 \pm 0.06 $   & $  0.84 \pm 0.07 $  
\enddata 
\tablenotetext{a}{The log of the best-fit offset to the simulations.
A positive value indicates that the prediction for the percentile is
too low. Errors are given by $\Delta\chi^2=1$.}
\tablenotetext{b}{These correspond to -0.5, 0.0, and +1.0 $\sigma$ in a lognormal distribution of $\tau$ about the median; see S03. There are 22, 41, and 35 degrees of freedom for percentiles 31, 50, and 84, in CIV/HI, and 24 d.o.f. for CIII/CIV.}
\tablenotetext{c}{For `C' models particles with $t_c<t_H$ are set to $T=2\times10^4$\,K. For `C1' models this is done if $t_c < 0.1t_H$.}
\tablenotetext{d}{For these models the simulation values were too low to obtain a reliable fit.}
\tablenotetext{e}{In the CIII/CIV results for these models, a yield given by the CIV/HI offset was used; these are marked by an `F' in Fig.~\ref{fig-civnocool}.}
\end{deluxetable} 
%\end{deluxetable*} 

\section{Conclusions}
\label{sec-conc}

A major goal for cosmological simulations is to develop a prescription
for feedback with which observations of the IGM and galaxies can be
simultaneously reproduced. We have compared in detail the statistics
of \CIV\ and \CIII\ absorption in a set of six high-quality $z\sim
3-4$ quasar spectra to that in simulated spectra drawn from two
tate-of-the-art cosmological SPH simulations with a wide range of UVB
models and two different prescriptions for feedback: the simulation of
Springel \& Hernquist (2003a), which predicts a numerically resolved
star formation history that is consistent with the observations and
the simulation of Theuns et al.\ (2002) which matches observations of
the \HI\ Ly$\alpha$ forest. We have come to the following conclusions:

\begin{itemize}

\item {The simulations predict far too little \CIV\ absorption unless
the UVB is extremely soft.}

\item {In all cases, the simulations predict far too small \CIII/\CIV\
ratios.}

\item The simulations predict that many of the heavy elements reside
  in gas that is metal-rich ($Z \ga Z_\odot$) and hot ($10^5\,
  \lesssim T < 10^7\,$K). Much of this gas should be able to cool via
  metal lines, which was not included in the simulations.  It will be
  important to accurately model cooling in future numerical
  simulations.

\item If a crude cooling prescription is applied, the \CIV\ absorption
increases significantly, but the predicted \CIII/\CIV\ ratio is still
far too low because the metals reside in gas that is too low density.
Cooling should increase the enriched gas density, and this can be
estimated in future simulations with metal-line cooling.
 
\item Independent of the cooling prescription, the metal distribution
  in the simulations is too inhomogeneous to match the observed
  distribution of $\tau_{\rm CIV}(\tau_{\rm HI})$.

\end{itemize}

Numerical simulations with strong outflows from $M_{\rm baryon}
\gtrsim 10^8\msol$ galaxies deposit most intergalactic metals in hot,
metal-rich bubbles that preferentially inhabit voids and comprise a
relatively small filling factor, which allows them to avoid overly
disrupting the Ly$\alpha$ forest (Theuns et al.\ 2002).  However,
these very attributes appear to prevent them from reproducing the
observations of absorption by heavy elements.

If metal cooling is efficient or if the gas has an unresolved
multiphase structure then these winds may account for some of the
observed enrichment; otherwise they would be largely hidden from
current observations.  In either case it appears that an additional
ingredient -- either in the form of another enrichment mechanism, or
higher-$z$ enrichment that is unresolved by the simulations -- is
needed.

\acknowledgements This work was supported by the W.M.~Keck foundation,
NSF grants PHY-0070928, AST 02-06299, and AST 03-07690, NASA ATP
grants NAG5-12140, NAG5-13292, and NAG5-13381, a PPARC Advanced
Fellowship, and Silicon Graphics/Cray Research.


\begin{thebibliography}{}

\bibitem[Aguirre et al.(2001)]{2001ApJ...560..599A} Aguirre, A.,
Hernquist, L., Schaye, J., Weinberg, D.~H., Katz, N., \& Gardner, J.\
2001, \apj, 560, 599

\bibitem[Paper(I)]{paper1} Aguirre, A., Schaye, J., \& Theuns, T.\
2002, \apj, 576, 1 (A02)

\bibitem[Aguirre et al.(2004)]{2004ApJ...602...38A} Aguirre, A., Schaye, 
J., Kim, T., Theuns, T., Rauch, M., \& Sargent, W.~L.~W.\ 2004, \apj, 602, 
38 (A04)

\bibitem[Anders \& Grevesse(1989)]{1989GeCoA..53..197A} Anders, E.~\& 
Grevesse, N.\ 1989, \gca, 53, 197 

\bibitem[Aracil, Petitjean, Pichon, \& Bergeron(2004)]{2004A&A...419..811A} 
Aracil, B., Petitjean, P., Pichon, C., \& Bergeron, J.\ 2004, \aap, 419, 
811 

\bibitem[Boksenberg et al.(2003)]{boksen} Boksenberg, A., Sargent, W.L.W., \& Rauch, M.\ 2003, \apjs, submitted; astro-ph/0307557

\bibitem[Cen, Nagamine, \& Ostriker(2004)]{2004astro.ph..7143C} Cen, R., 
Nagamine, K., \& Ostriker, J.~P.\ 2004, astro-ph/0407143

\bibitem[Cowie \& Songaila(1998)]{1998Natur.394...44C} Cowie, L. L. \& 
Songaila, A. 1998, \nat, 394, 44 

\bibitem[Cowie et al.(1995)]{1995AJ....109.1522C} Cowie,
L.~L., Songaila, A., Kim, T., \& Hu, E.~M.\ 1995, \aj, 109, 1522 

\bibitem[Ellison et al.(2000)]{2000AJ....120.1175E} Ellison, S.~L., Songaila, A., Schaye, J., \& Pettini, M.\ 2000, \aj, 120, 1175 

\bibitem[Furlanetto \& Loeb(2003)]{2003ApJ...588...18F} Furlanetto, 
S.~R.~\& Loeb, A.\ 2003, \apj, 588, 18

\bibitem[Gnedin(1998)]{1998MNRAS.294..407G} Gnedin, N.~Y.\ 1998, \mnras, 
294, 407 

\bibitem[Haardt \& Madau(2001)]{haardt01:cuba}
Haardt, F.~\& Madau, P. 2001, astro-ph/0106018

\bibitem[Hernquist \& Springel(2003)]{2003MNRAS.341.1253H} Hernquist, L.~\& 
Springel, V.\ 2003, \mnras, 341, 1253 

\bibitem[Kay, Pearce, Frenk, \& Jenkins(2002)]{2002MNRAS.330..113K} Kay, 
S.~T., Pearce, F.~R., Frenk, C.~S., \& Jenkins, A.\ 2002, \mnras, 330, 113 

\bibitem[Paper(II)]{paper2} Schaye, J., Aguirre, A., Kim, T., Theuns,
T., Rauch, M., \& Sargent, W.L.W.\ 2003, \apj, 596, 768 (S03)

\bibitem[Scannapieco, Ferrara, \& Madau(2002)]{2002ApJ...574..590S} 
Scannapieco, E., Ferrara, A., \& Madau, P.\ 2002, \apj, 574, 590 


\bibitem[Simcoe, Sargent, \& Rauch(2004)]{2004ApJ...606...92S} Simcoe, 
R.~A., Sargent, W.~L.~W., \& Rauch, M.\ 2004, \apj, 606, 92

\bibitem[Shapley, Steidel, Pettini, \& 
Adelberger(2003)]{2003ApJ...588...65S} Shapley, A.~E., Steidel, C.~C., 
Pettini, M., \& Adelberger, K.~L.\ 2003, \apj, 588, 65

\bibitem[Songaila \& Cowie (1996)]{1996AJ....112..335S} Songaila, A.  \&
Cowie, L. L. 1996, \aj, 112, 335

\bibitem[Springel \& Hernquist(2003)]{2003MNRAS.339..312S} Springel, V.~\& 
Hernquist, L.\ 2003a, \mnras, 339, 312 (SH03)

\bibitem[Springel \& Hernquist(2003)]{2003MNRAS.339..289S} Springel, V.~\& 
Hernquist, L.\ 2003b, \mnras, 339, 289 

\bibitem[Sutherland \& Dopita(1993)]{1993ApJS...88..253S} Sutherland, 
R.~S.~\& Dopita, M.~A.\ 1993, \apjs, 88, 253

\bibitem[Theuns et al.(2002)]{2002ApJ...578L...5T} Theuns, T., Viel, M., 
Kay, S., Schaye, J., Carswell, R.~F., \& Tzanavaris, P.\ 2002, \apjl, 578, 
L5 (T02)


\end{thebibliography}
\end{document}